\documentclass[preprint,12pt]{elsarticle}

\usepackage{amssymb}
\usepackage{amsmath}
\usepackage{gensymb}
\graphicspath{ {./images/} }
\usepackage{caption}
\usepackage{subcaption}
\usepackage{algorithm}
\usepackage{algorithmicx}
\usepackage{algpseudocode}
\journal{Aerospace Science and Technology}

\begin{document}

\begin{frontmatter}



\title{A novel Artificial Neural Network-based streamline tracing strategy applied to hypersonic waverider design }


\author[inst1]{Anagha G. Rao}
\author[inst1]{Umesh Siddharth}
\author[inst1]{Srisha M. V. Rao$^*$}

\affiliation[inst1]{organization={Department of Aerospace Engineering, Indian Institute of Science},%Department and Organization
            city={Bengaluru},
            postcode={560012}, 
            state={Karnataka},
            country={India}}

\begin{abstract}
Streamline tracing in conical hypersonic flows is essential for designing high-performance waverider and intake. Conventionally, the streamline equations are solved after obtaining the velocity field from the solution of the axisymmetric conical flow field. The hypersonic waverider shape is generated from the base conical flow field by repeatedly applying the streamline tracing approach along several planes. When exploring the design space for optimization of the waverider, streamline tracing can be computationally expensive. We provide a novel strategy where first the Taylor-Maccoll equations for the inviscid axisymmetric conical flowfield and the streamlines from the shock are solved for a wide range of cone angle and Mach number conditions resulting in an extensive database. The streamlines are parametrized by a third-order polynomial, and an Artificial Neural Network (ANN) is trained to predict the coefficients of the polynomial for arbitrary inputs of Mach number, cone angle, and streamline originating location on the shock . We apply this strategy to design a cone derived waverider and compare the geometry obtained with the standard conical waverider design method and the simplified waverider design method. The ANN technique is highly accurate, with a difference of 0.68\% with the standard in the coordinates of the waverider. RANS computations show that the ANN derived waverider does not indicate severe flow spillage at the leading edge, which is observed in the waverider generated from the simplified method. The new ANN-based approach is 20 times faster than the conventional method.
\end{abstract}

\begin{graphicalabstract}
\includegraphics[width=\textwidth]{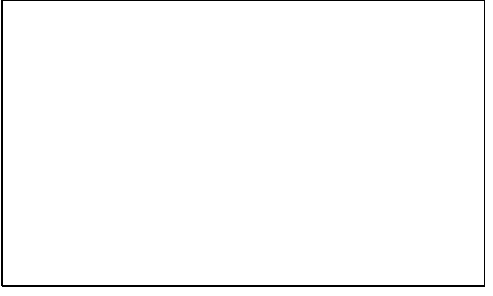}
\end{graphicalabstract}

\begin{highlights}
\item Streamline tracing in axisymmetric flowfields is crucial for hypersonic vehicle design. 
\item The standard technique requires the solution of a system of differential equations. 
\item We have developed a novel computationally efficient ANN-based streamline tracing method. 
\item The new ANN technique is applied to conical waverider design. 
\item The ANN technique has an error of 0.68\% and is 20 times faster than the standard technique. 
\end{highlights}


\end{frontmatter}


\section{Introduction}
\label{intro}
Contemporary mission requirements of long-range manoeuvrable hypersonic vehicles for space, military and transport applications demand innovations in vehicle design. Unique aerothermodynamic challenges in the hypersonic flow regime necessitate a fully integrated hypersonic vehicle geometry \cite{Anderson2006,Bertin1994}. The hypersonic waverider whose leading edges ride on the shock wave, thereby encapsulating a region of high pressure, provides a significantly high lift to drag ratio essential for modern-day applications. Under the integrated vehicle design philosophy, large potions of the vehicle forebody are contoured as an intake to efficiently deliver compressed air to the air-breathing propulsion system. Streamline tracing forms an indispensable technique in designing high-performance intakes and waveriders.
\par 
Hypersonic intakes derived from axis-symmetric converging flow field known as Busemann flow field have been shown to give high-efficiency compression \cite{VanWie1992,Billig2000}. In a detailed chapter, S Molder \cite{Molder2019} has discussed the benefits and procedures to calculate the Busemann intake. Streamline tracing is an essential step in the procedure. Further, the streamline tracing technique can be exploited to obtain intake shapes that suitably converge from the entry to an exit of the shape desired at the combustor resulting in different arrangements such as the sugar scoop-shaped intake. The startability of the sugar scoop-shaped intake and its control by means of a sliding door has been experimentally tested at the Virginia Supersonic Windtunnel at Mach number of 4 \cite{Jacobsen2006}.
\begin{figure}[htb]
\includegraphics[width=\textwidth]{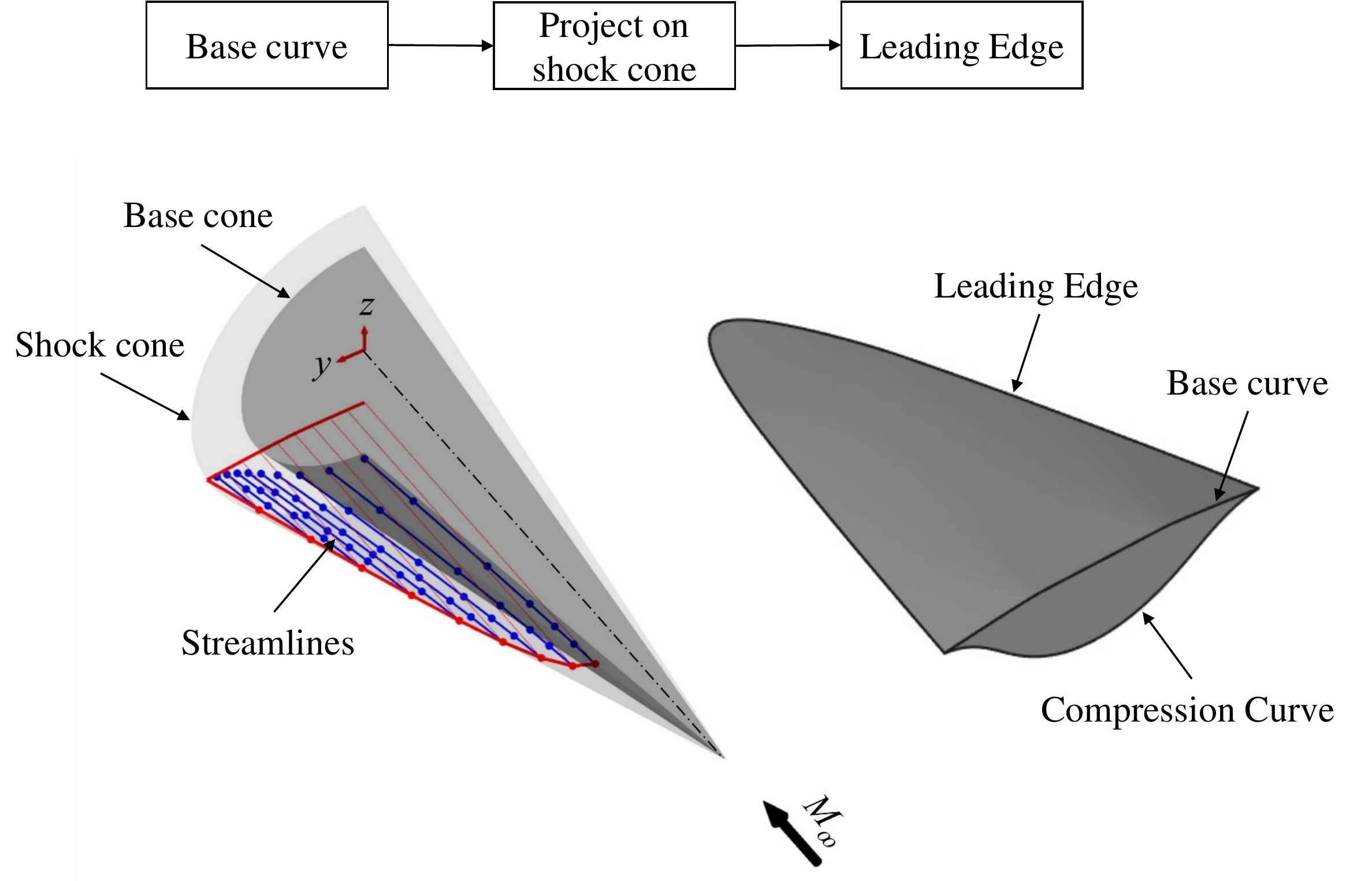}
\caption{Pictorial illustration of waverider geometry generation and waverider terminologies}
\label{fig01}
\end{figure}
\par
Kuchemann \cite{Kuchemann1965} in an overarching study emphasized the peculiar challenges of designing hypersonic vehicles in comparison to subsonic or supersonic vehicles and, in this context, highlighted the need for lifting bodies such as the waverider and detailed several waverider configurations. The waverider concept was first introduced by Nonweiler \cite{Nonweiler1959}, where the leading edges of the vehicle ride the shock wave while other surfaces are suitably recessed and showcased the caret shaped waverider. Typically, the waverider consists of a freestream surface parallel to the freestream and a compression surface that forms the underbelly of the vehicle. The leading edge, which is the intersection of the freestream surface and the compression surface, rides the shock wave of the body, encapsulating a high pressure region under the compression surface, thereby generating a significantly higher lift. The base surface comprises the freestream surface (base) curve and the compression surface curve in a plane perpendicular to the freestream. Waverider shapes are generated from a basic flow field by projecting the base curve onto the shock and using streamline tracing to generate the compression surface as depicted pictorially in Figure \ref{fig01}. The wedge based flow field generates an oblique shock which is easily computed. The caret shaped waverider and the power-law waverider are typical examples of wedge-based waveriders \cite{Starkey1998}. Besides the lift to drag ratio, L/D, the volume contained within the waverider is important from the payload perspective. Hence, a good volumetric efficiency is essential. Waveriders can be generated from conical flow fields, which can be computed using the Taylor-Maccoll equations. In some cases, generic three-dimensional shapes have also been used as the basic flow field. Recently, state of the art in waverider design methodologies have been comprehensively described in a review article\cite{Ding2017}.
\par
Wedge derived waveriders formed the basis of developing the conceptual design of a complete air-breathing propulsion hypersonic vehicle by Ferguson et al. \cite{Ferguson2008}. A modular design where multiple waverider configurations were combined to generate a star-shaped aerodynamic surface was described. However, requirements of larger volumetric efficiencies led to a shift toward cone based waverider designs in a later study \cite{Ferguson2015}. Jones et al. \cite{Jones1968} have described the generation of simple cone base waveriders from axisymmetric right circular cone flow field and have classified two kinds of waveriders, one where the apex of the waverider coincides with the axis of the cone and the second where it is offset from the cone axis. A high degree of design flexibility is required to meet the several competing objectives of hypersonic vehicles, for example, the need for bluntness at the leading edges to avoid severe aerodynamic heating. Therefore, many improvements to the design process continue to be made even today. A generalized inverse design method where the shock wave shape is given and axis-symmetric flow fields are utilized to generate the given shock curve by assuming local axis-symmetric flow field along osculating planes significantly enhance the design space for waveriders \cite{Sobieczky1990}. The osculating cone method is a particular case of the general method which has been used extensively in recent times. In a further development, the requirement of the same axis-symmetric shape for the flow field at each osculating plane has been relaxed, leading to the osculating flow field method for waverider design \cite{Rodi2005}. Ding et al. \cite{Ding2015,Ding2013} have proposed a simplification to the streamline tracing method of the conical flow field, and the resulting waverider has shown an increase in volumetric efficiency. However, flow spillage has been observed at the leading edge. The method of computing the axis-symmetric flow field plays a role in determining the rapidity and accuracy of waverider design and parametric analysis. Analytical solutions from hypersonic small disturbance theory enable rapid assessment of a large design space, but their applicability is restricted to small cone angles \cite{Rasmussen1980,Mangin2006}. In the majority of the studies, Taylor-Maccoll relations are solved to obtain the axis-symmetric flow field. High fidelity Euler solutions allow the generation of waveriders from a more general class of conical shapes which need not be restricted to right circular cones, but they come with high computing costs and time \cite{Cui2007}. Chen et al. \cite{Chen2019} have improvised the osculating cone technique by adjusting the radius of the cone to achieve a target volume of the waverider. Attempts have been made to make waverider design suitable to a the wide range of Mach numbers by providing a Mach number profile at each osculating plane along the prescribed shock curve \cite{Liu2019}, or by discretizing the leading edge curve for different Mach numbers \cite{Li2018}. The streamline tracing technique has been extended to waverider generation from a wide variety of basic flows, including wedge-cone combination \cite{Takashima1994}, power-law blunt bodies \cite{Mangin2006}, Von Karman ogive \cite{Ding2015b}, to name a few. In some instances, viscous effects have been considered by using integral boundary layer equations after solving the inviscid flow field over the basic shape before finally deriving the waverider shape \cite{CORDA1988,Mangin2006}. Integration of the intake with the waverider remains a key challenge. Novel strategies have been developed: in one study, a combined axisymmetric base flow consisting of a cone with the intake cowl has been taken as the basic flow field \cite{Ding2015a}, and in another, a suitable merging of inward turning and outward turning axis-symmetric flow field has been achieved \cite{Li2020}. Thus, with multiple competing objectives to be satisfied, the waverider design principles continue to evolve. Generally, the newly developed waverider design method is compared with existing methods using inviscid, and RANS CFD computations \cite{Ding2015,Chen2019,Ding2015b}. Rapid computation of waverider geometry is crucial to explore the design space for optimization purposes, which requires efficient parametrization and predictive capabilities.
\par
\begin{figure}[htb]
\includegraphics[width=\textwidth]{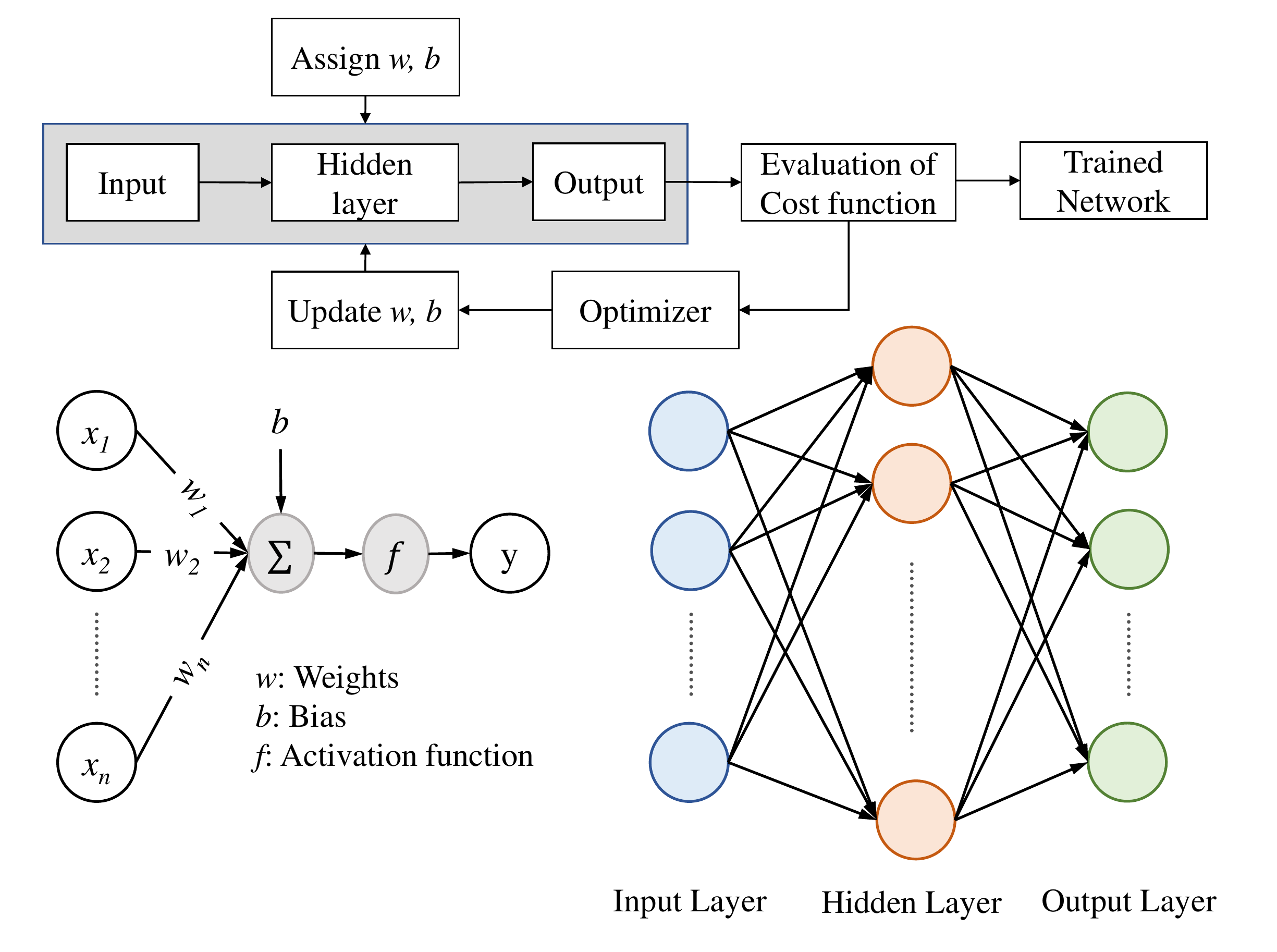}
\caption{Typical architecture of ANN and the workflow for training the ANN}
\label{fig02}
\end{figure}

Modern-day data-driven Artificial Intelligence (AI) tools have revolutionized data analysis and prediction capabilities. MATLAB provides several algorithms for Machine Learning (ML), Artificial Neural Networks (ANN), and AI, which are elaborately described in the reference \cite{Kim2017}. The ANN consists of interconnected layers of nodes that function analogous to biological neurons. They are known to excellently approximate functions between multiple inputs to multiple outputs of heterogeneous categories. The ANN learns from a known database presented to it by calculating the weights of the interconnections between nodes of different layers by minimizing a suitably defined error function. Post learning, the ANN can be used as a predictive tool. The general architecture of an ANN consists of one input layer, one output layer, and several hidden layers as depicted in Figure \ref{fig02}. The overall workflow for training the ANN is also represented in the figure. The ANN has been deployed in the solution of differential equations \cite{Lagaris1998}, and the numerical thermochemical computations for hypersonic flows \cite{Ozbenli2020}. Kutz \cite{Kutz2017} briefly describes the rapid rise in the use of Deep Neural Networks (DNN) for turbulence modelling and prediction. Miyanwala and Jaiman have successfully used DNN to solve the Navier-Stokes equation for the unsteady vortex shedding problem behind a cylinder \cite{Miyanawala2017}. Aerodynamic shape optimization requires large scale computations of several parameters, which is computationally expensive and time-consuming. ANN has been used as a surrogate model by learning from a limited dataset to predict a large design space as in the cases for the airfoil and wing geometry \cite{Li2020a,Tao2019,Hui2021}. Flow field reconstruction from experimental and high fidelity CFD data has also gained significantly from deep learning techniques, for example, in the prediction of flow field over airfoils \cite{Sekar2019}, and the velocity field in scramjet isolator \cite{Kong2021}. Fujio and Ogawa \cite{Fujio2021} utilize a surrogate assisted evolutionary optimization framework in optimizing a streamline traced sugar-scoop type intake. A generalized ANN-based model of the supersonic ejector has been developed by the author's group, which has the capability to predict performance for different working fluids \cite{Gupta2022}.
\par
Hypersonic waveriders are preferred shapes for long-range manoeuvrable hypersonic vehicles. Predominantly axis-symmetric cone flow fields solved by the Taylor-Maccoll equations are utilized to design waveriders. For a single waverider shape, several streamline tracing calculations have to be carried out to define the compression surface, which involves marching solutions to the streamline differential equations. When considering the optimization of the waverider for multiple objectives such as high aerodynamic efficiency, volumetric efficiency, and low aerodynamic heating, effective parametrization and computation of the waverider geometry over a large design space must be accomplished in the shortest duration. However, solving differential equations is computationally expensive and time-consuming. ANN has been effectively used as an approximation to function in solution of differential equations. However, the application of ANN for streamline tracing leading to rapid waverider shape generation has not been explored in the literature which motivated the current investigation. Firstly, a parametrized form of the streamline equations in terms of a polynomial fit with high prediction accuracy is obtained from the solution of Taylor-Maccoll equations for a wide range of Mach numbers. The coefficients of the polynomial fit are then predicted using a ANN, which has been trained using the extensive database generated. Waveriders generated from the standard cone method, simplified cone method, and the proposed ANN method are compared for shapes and final flow fields by CFD solutions of RANS equations. We have successfully achieved highly accurate waverider geometries using the ANN method with significant reduction in computational time.
\section{Methodology}\label{method}
\subsection{Cone derived waverider design}\label{design}
\subsubsection{Taylor Maccoll solution}

\begin{figure}[htb]
\includegraphics[width=\textwidth]{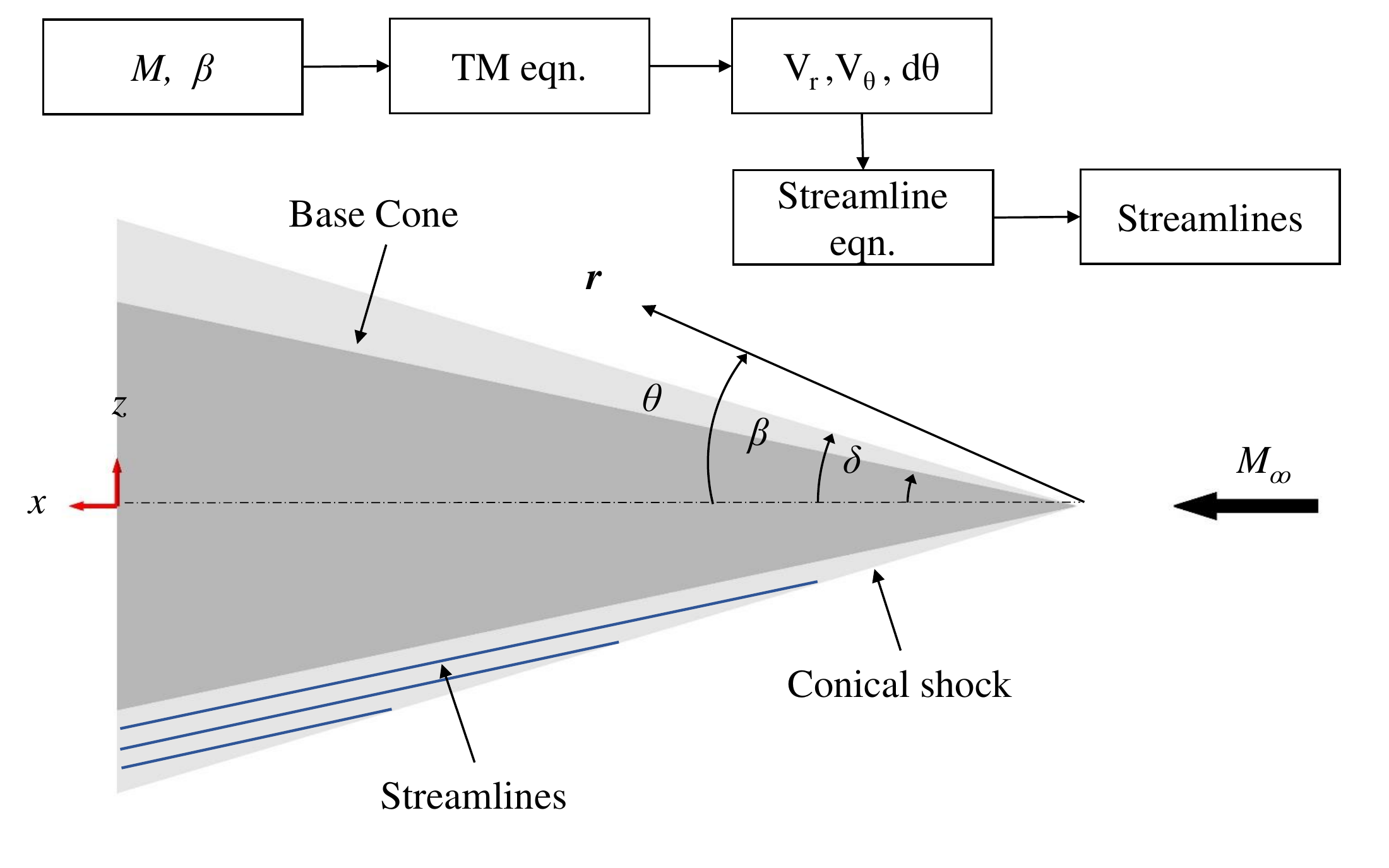}
\caption{An illustration of streamlines traced from conical shock, $r$ and $\theta$ refer to radial and azimuthal coordinates, $\beta$ is the shock angle and $\delta$ is the flow deflection/cone semi-vertex angle }
\label{fig03}
\end{figure}

The inviscid axis-symmetric conical flow field over a right circular cone in supersonic flow is first solved using the Taylor-Maccoll (TM) equation given in Equation \ref{eqn01}, where $V_r$ and $V_\theta$ are the velocity components in the radial and azimuthal directions. Equation \ref{eqn02} gives relation for $V_\theta$ in terms of $V_r$, where $V'$ is defined by Equation \ref{eqn03}. The boundary conditions are the oblique shock jump conditions at the shock and flow tangency condition at the wall. In this work, the TM equation is solved using MATLAB with a fourth-order Runge-Kutta function to obtain the flow field between the shock and the cone wall as illustrated in Figure \ref{fig03}.

\begin{equation}
\frac{\gamma-1}{2}\left[ 1-V'_r{}^{2}- \left( \frac{dV'_r}{d\theta} \right)^2 \right] \left[ 2V'_r-\frac{dV'_r}{d\theta}\cot\theta+ \frac{d^2V'_r}{d\theta^2} \right]-
\frac{dV'_r}{d\theta}\left[V'_r\frac{dV'_r}{d\theta}+\frac{dV'_r}{d\theta}\frac{d^2V'_r}{d\theta^2} \right]=0 \label{eqn01}
\end{equation}
\begin{equation}
    V'_\theta=\frac{d V'_r}{d \theta}\label{eqn02}
\end{equation}
\begin{equation}
    V'=\frac{V}{V_{max}}=\left[\frac{2}{\left(\gamma-1\right)M^2}+1\right]^{-0.5}\label{eqn03}
\end{equation}
\subsubsection{Streamline Tracing}

Following the solution of the TM equation, the streamline differential equation represented in Equation \ref{eqn04}  can be evaluated by numerically marching from points located on the shock to obtain the streamline points. A schematic representation of the streamlines traced from the right circular cone shock is depicted in Figure \ref{fig03}. The embedded flowchart represents the steps to obtain the streamlines from inputs ($M$,$\beta$) to the streamline points.  
\begin{equation}
\frac{dr}{V_r}=\frac{rd\theta}{V_\theta}\label{eqn04}
\end{equation}    

\subsubsection{Generation of waverider geometry}

\begin{figure}[htb]
\centering
\includegraphics[width=0.7\textwidth]{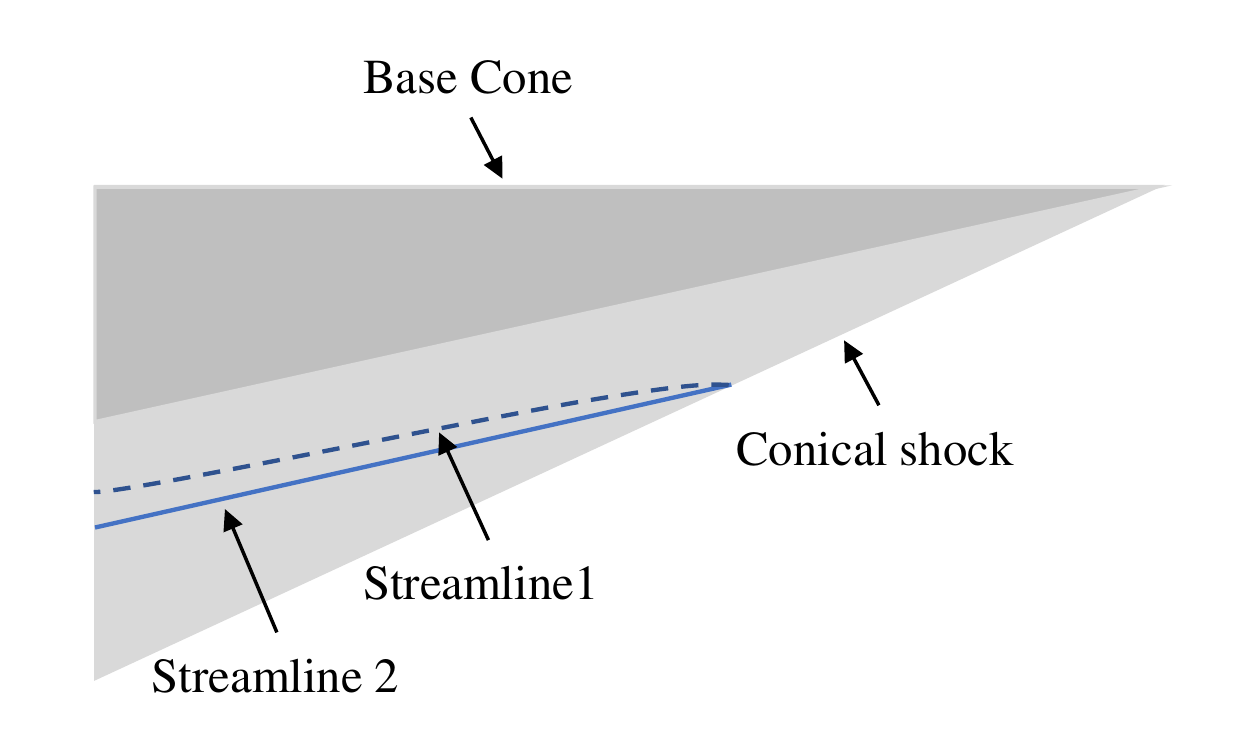}
\caption{Schematic of streamlines generated using 1. Standard and 2. Simplified method}
\label{fig04}
\end{figure}

The base curve has to be specified to obtain the waverider geometry using the cone derived waverider method. The base curve is projected onto the shock by generators parallel to the freestream to obtain the leading edge. Thereafter, streamline tracing is carried out from leading edge locations on the shock back to the base plane to generate the compression surface. Two different streamline tracing techniques have been followed : in the standard cone derived waverider technique the streamline equation (Equation \ref{eqn04}) is solved, whereas in the simplified technique, streamlines are taken tangential to the base cone surface to easily obtain the compression surface points. Algorithm \ref{stdconewaveriderslgo} details the steps for both the standard and simplified methods of cone derived waverider geometry generation.  Figure \ref{fig04} pictorially represents the differences in the streamlines generated using standard technique and the simplified technique. 

\begin{algorithm}[htb]
\caption{Cone derived waverider design procedure}
\begin{algorithmic}
\State Input: Mach number, cone angle $\theta$ /cone shock angle $\beta$
\State Input: Freestream curve at the base

    \If {M and $\theta$ given}
    \State Calculate $\beta$ using cone shock theory (solution of TM equations)
    \Else
    \State Use M and $\beta$ 
    \EndIf
\State Solve TM equations to obtain $V_r$ and $V_{\theta}$

\State Step 1: Use straight lines parallel to the freestream to project the base curve onto the shock to produce the leading edge curve
    \If {standard cone derived waverider}
    
    \State Step 2: Starting from discrete locations on the leading edge curve carry out streamline tracing by solving Equation \ref{eqn03}  marching toward the base plane
    \EndIf
    \If {simplified cone derived waverider}
    \State Step 2: Starting from discrete locations on the leading edge curve carry out streamline tracing by considering straight lines tangential to the cone surface
    \EndIf
\State Step 3: Collection of such streamline traced curves form the compression surface

\end{algorithmic}
\label{stdconewaveriderslgo}
\end{algorithm}
   
\subsubsection{Base Curve}\label{Base_curve}
\begin{figure}[htb]
\centering
\includegraphics[width=0.5\textwidth]{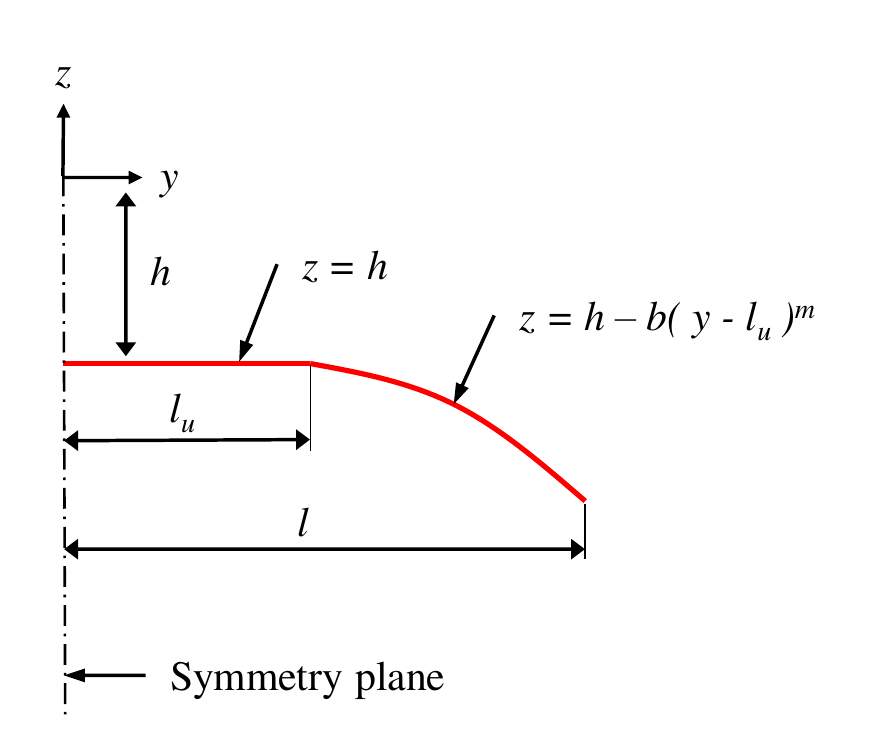}
\caption{Base curve}
\label{fig05}
\end{figure}

The base curve equation employed for waverider design is presented about the symmetry plane in Figure \ref{fig05}. $l_u$ forms the length of flat portion of the curve while $l$ is width of the waverider. The curved portion of the curve is given by the power law equation. The value of b can be determined for given $l_u$, $l$ and $h$.

\subsection{The novel ANN based streamline strategy}
\begin{figure}[htb]
\centering
\includegraphics[width=\textwidth]{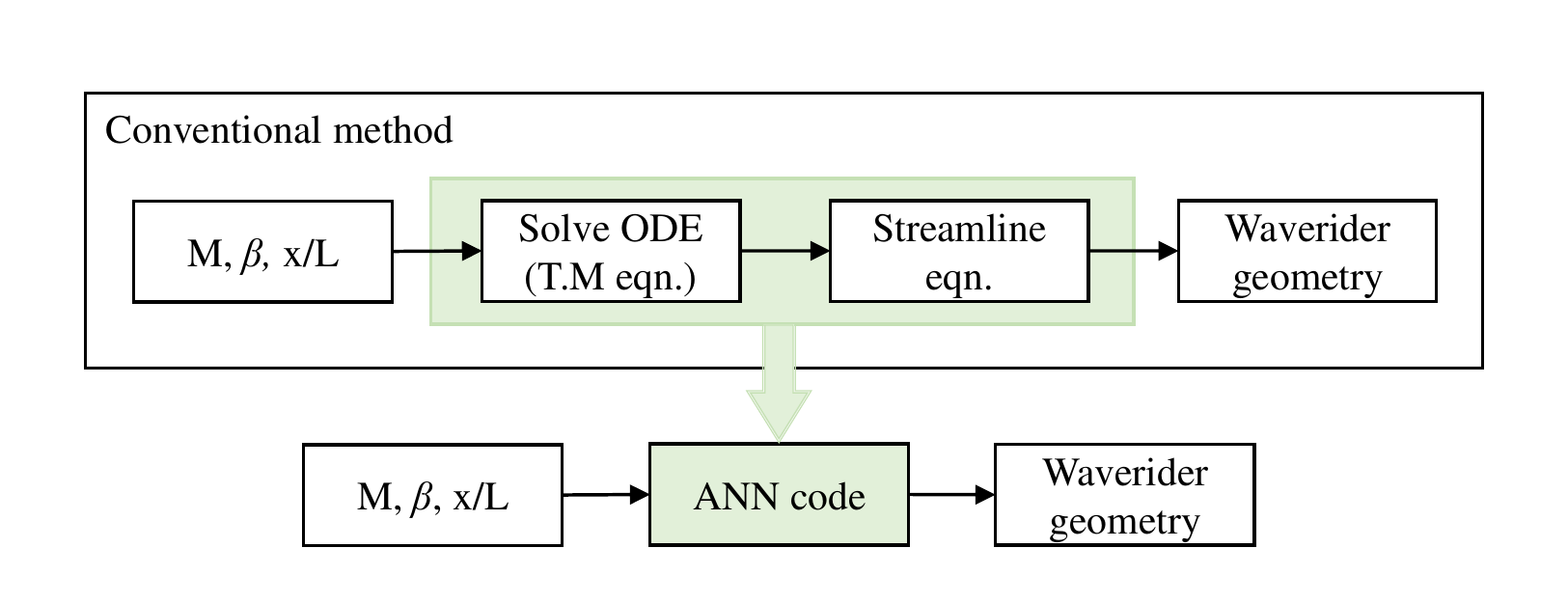}
\caption{A comparison of the conventional and the novel ANN strategy for waverider design}
\label{fig06}
\end{figure}
The standard design procedure requires the solution of TM and streamline differential equations as described in Section \ref{design} which is computationally expensive. A novel ANN based strategy is introduced in this work to trace the streamlines and subsequently generate the waverider geometry. Figure \ref{fig06} compares the standard procedure represented as a flowchart with the novel ANN based technique. Two blocks - solving TM equation and Streamline equation, in the conventional technique are replaced by a single block involving utilization of a trained ANN to predict the streamlines directly from the given inputs - Mach Number ($M$), shock angle ($\beta$), and the location along the shock ($x_s/L$).
\par The generation of the trained ANN involves parameterization of the streamline curve using a polynomial fit followed by training the ANN for the coefficients of the polynomial for given inputs of Mach number ($M$), shock angle ($\beta$) and streamline starting location on the shock ($x_s/L$).
\begin{equation}\label{eqn05}
    Y=C_0X^n+C_1X^{n-1}+C_2X^{n-2}+....+C_n
\end{equation}

\par Firstly, a large dataset of parameterised streamline curve coefficients is generated by solving the TM equation and the Streamline equation. Equation \ref{eqn05} represents the parameterised form of  the streamline curve approximated as a polynomial where $X=x/L$ , $Y=y/L$, $(x,y)$ are points on the streamline, and $L$ is the length of the cone. The coefficients of the polynomial curve fit, $C_n=f(M,\beta,x_s/L)$, are unique functions of only $M$, $\beta$ and $X_s$. $Y_s$ is not independent since $Y_s=X_s tan(\beta)$.
\begin{figure}[htb]
\centering
\includegraphics[width=0.7\textwidth]{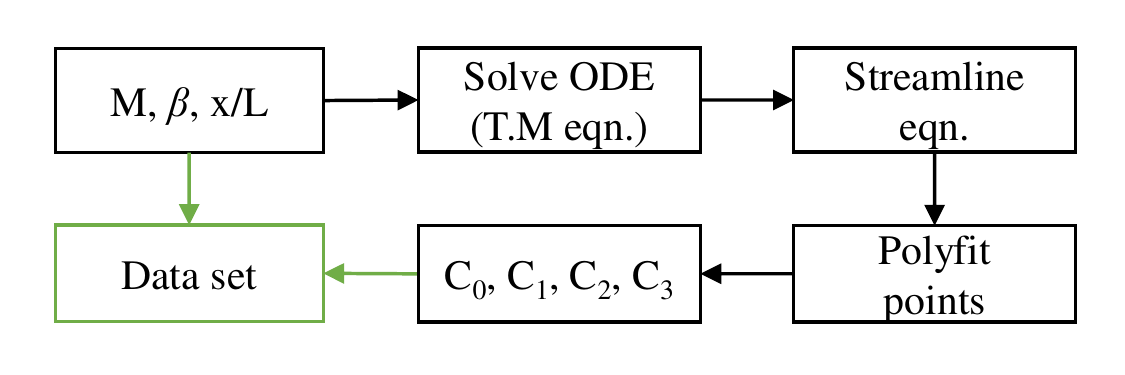}
\caption{A flowchart describing generation of data set }
\label{fig07}
\end{figure}

\begin{figure}[htb]
\centering
\includegraphics[width=\textwidth]{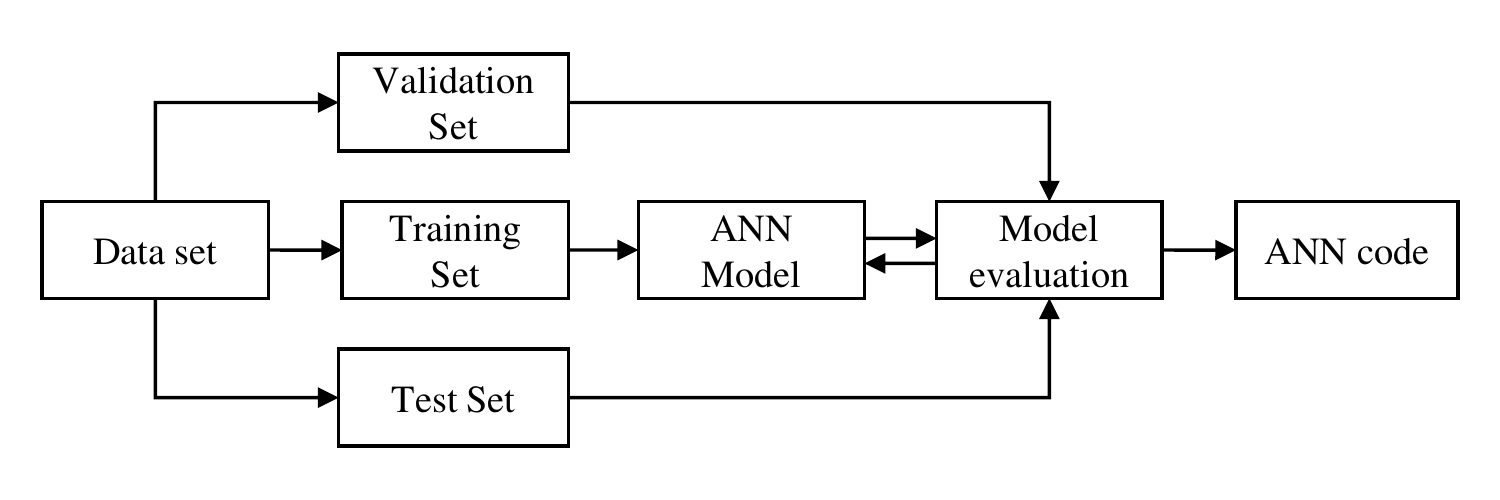}
\caption{Flowchart representing the ANN training process}
\label{fig08}
\end{figure}
For a given combination of $(M,\beta, X_s)$, the TM equation and the streamline equations are solved to obtain the points on the streamline curve. A polynomial regression fit is carried out to obtain the polynomial expression for the streamline with corresponding coefficients $(C_0,C_1,C_2, ... ,C_n)$. The process is repeated for discrete but a wide range of $(M,\beta, X_s)$ combinations within limits of the valid conical shock solutions. This results in the dataset which is further used for training the ANN. The process is represented as a flowchart in Figure \ref{fig07}. 
\par Different orders of the polynomial are tested for accuracy and compactness of the resulting dataset. The third order polynomial is found to be the best in both the measures. More details on the variation of errors with degree of polynomial is described in Section \ref{R&D}.  Moreover, the variation in fitness of the polynomial due to change in M and $\beta$ are found to be minimal.

\par A fully connected feed-forward Artificial Neural Network with tanh activation function is trained for $(M,\beta,X_s)$ as inputs and the coefficients of the polynomial fit $(C_0,C_1,C_2, ..., C_n)$ as outputs. $tanh$ function is used as the activation function at each node in the network. The entire dataset is divided into training, testing and validation sets. The training and testing datasets are used to train the ANN and evaluate its accuracy by means of back-propagation and adjustment of wieghts and biases at each node for error minimization. The validation set is finally used to obtain the global performance in terms of accuracy of prediction after obtaining the ANN model. Ultimately, the validated ANN model is utilized in the waverider design process. The steps followed in obtaining the ANN streamline tracing model is outlined in the flowchart represented in Figure \ref{fig08}.
\par A study is carried out by varying the number of hidden layers and the number of neurons in each hidden layer to obtain a compact network architecture with high accuracy. The influence of choosing one among three different optimization algorithms, viz. Levenberg Marquardt, Bayesian Regularisation, and Scaled Conjugate Gradient is also evaluated. The finalised ANN model consists of a single hidden layer network with 32 neurons and Bayesian Regularisation optimizer. 
\par The novel ANN streamline tracing model is then integrated with the cone derived waverider approach, where the streamline curve equation arising from specific points on the leading edge is directly predicted by the ANN model. The point cloud representing the waverider surface are generated from the known curves of the freestream surface and the compression surface. The waverider geometry obtained from the point cloud is taken up for flow analysis using ANSYS CFD tools.

\section{CFD}
\begin{figure}[htb]
\centering
\includegraphics[width=\textwidth]{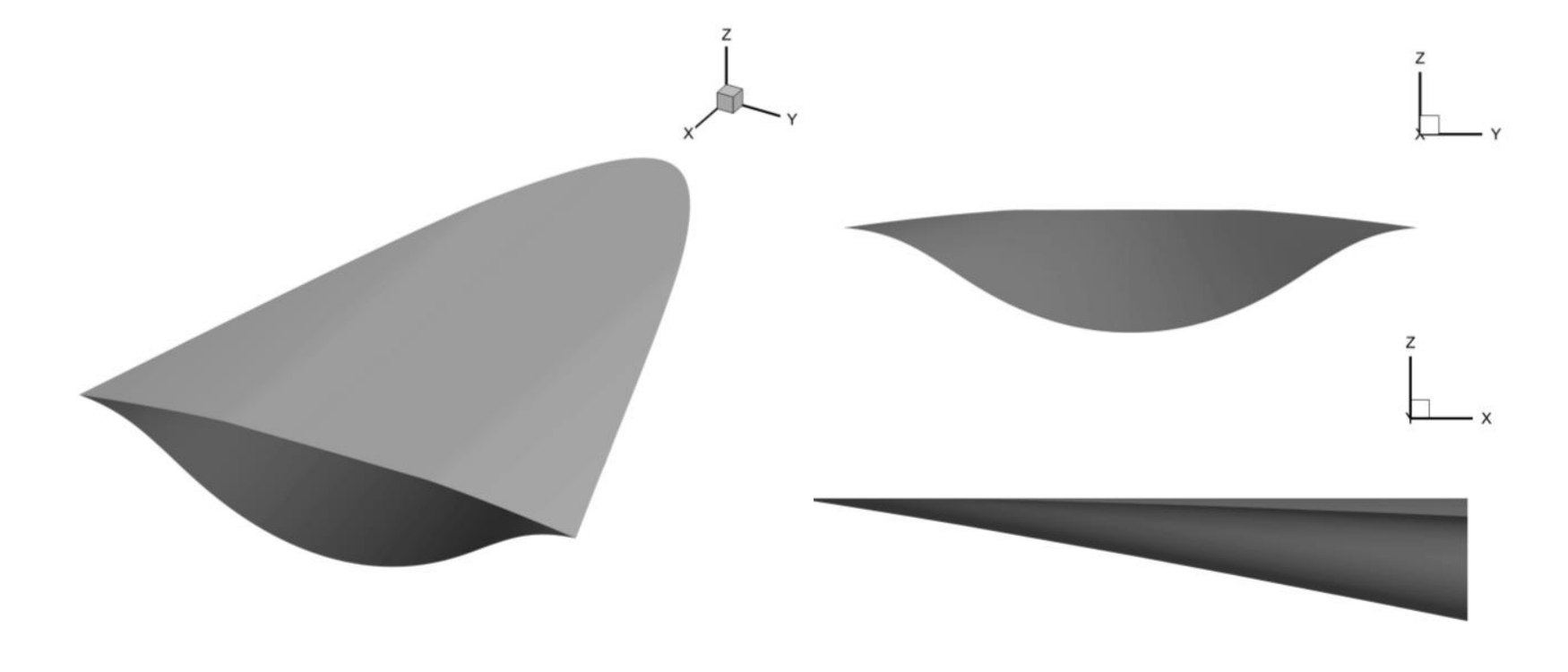}
\caption{Waverider}
\label{fig09}
\end{figure}
\subsection{Mesh}
\begin{figure}[htb]
  \begin{subfigure}[t]{0.5\textwidth}
    \includegraphics[width=\textwidth]{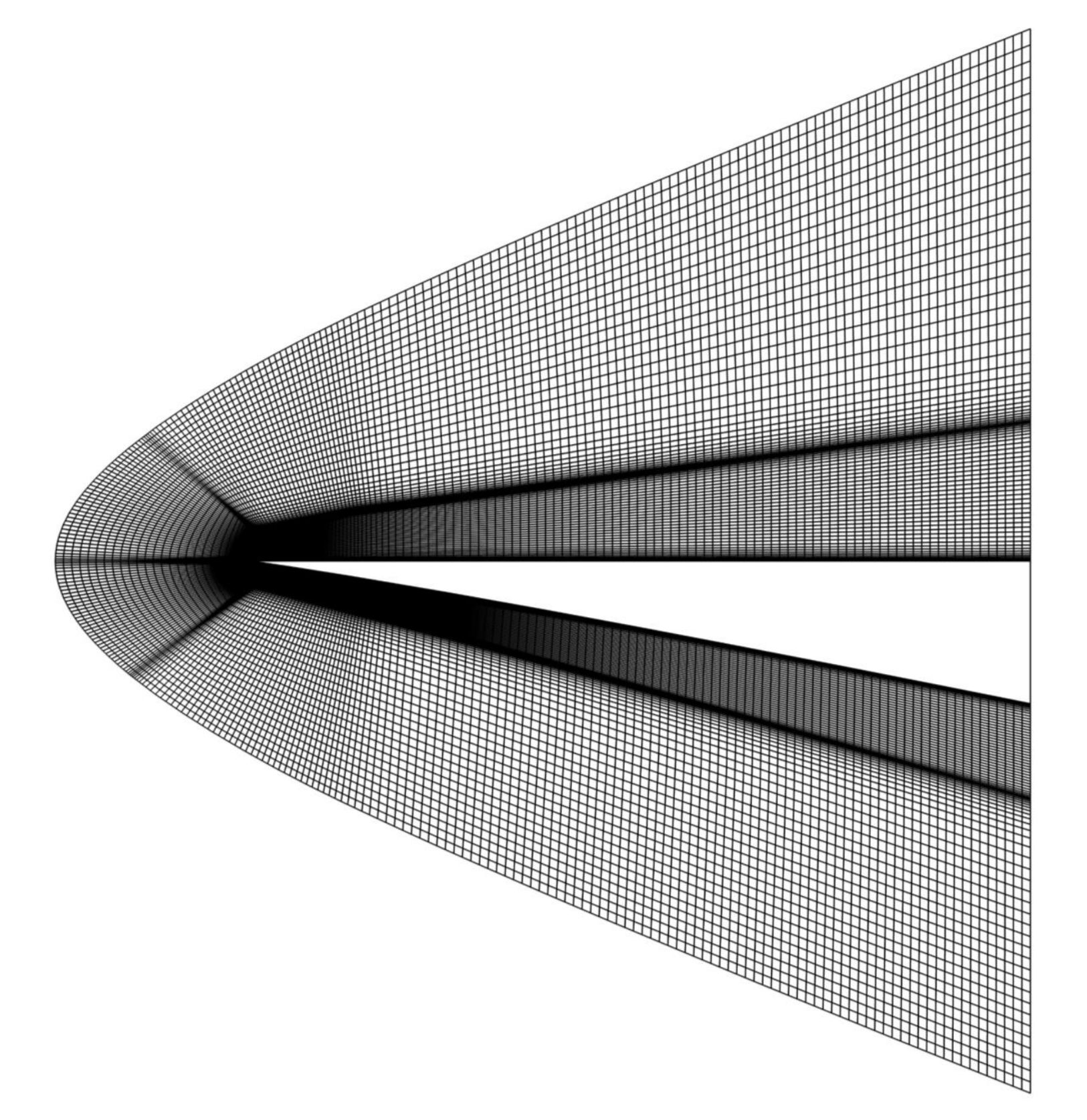}
    \caption{Symmetry plane}
    \label{fig10a}
  \end{subfigure}
  \hfill
  \begin{subfigure}[t]{0.35\textwidth}
    \includegraphics[width=\textwidth]{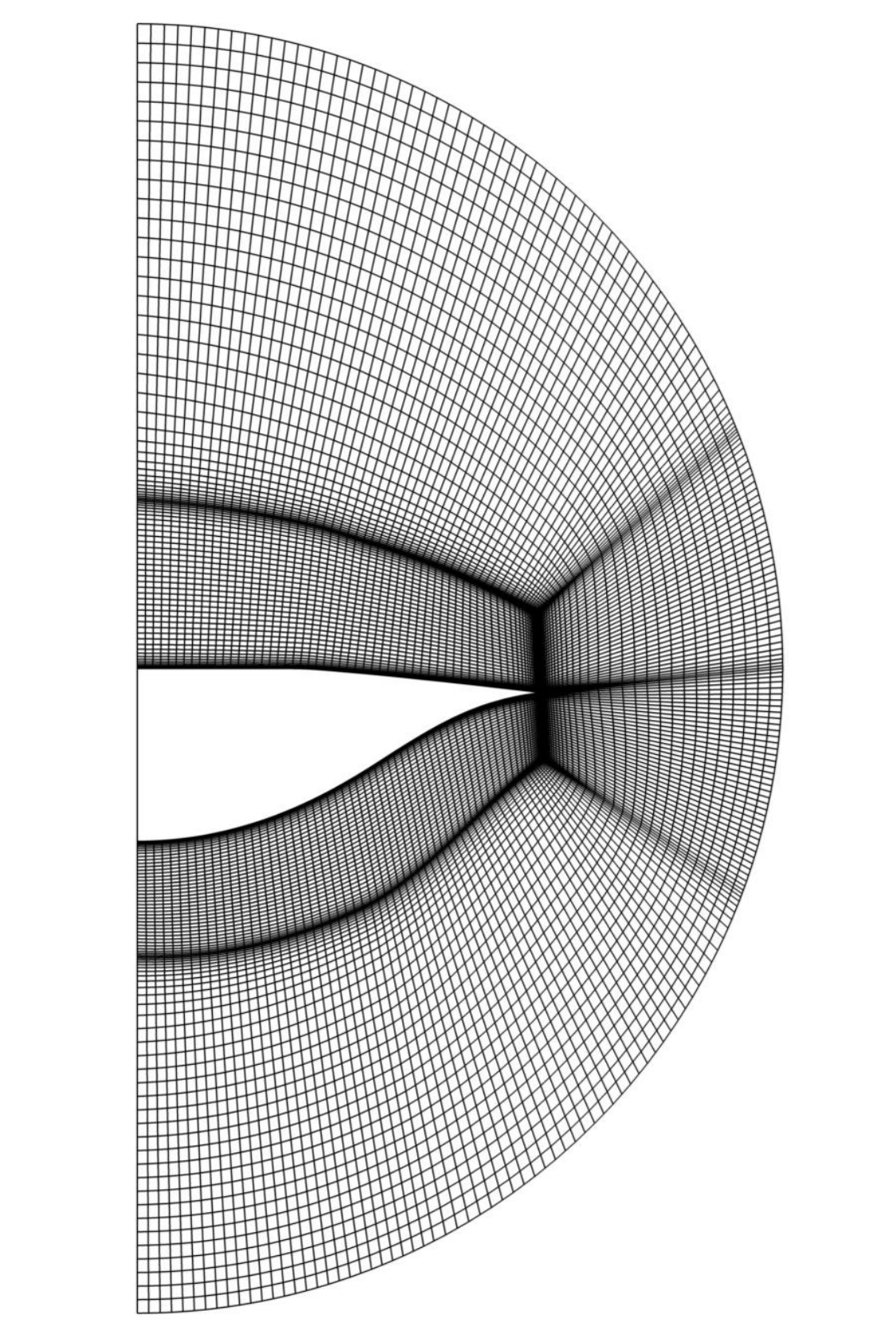}
    \caption{Base plane}
    \label{fig10b}
  \end{subfigure}
  \caption{Schematic of mesh employed for the flowfield}
  \label{fig10}
\end{figure}

A comparative study is carried out on three configurations of cone derived waveriders modelled by the standard, the simplified and the novel ANN based methodology. The models are designed for base cone angle of 12\degree and base curve as presented in Section \ref{Base_curve}. Three views of the standard waverider model are presented in Figure \ref{fig09}. Furthermore, the models are meshed in the software ICEM CFD. Due to symmetric nature of the geometry, one half of the model is considered for simulation. Figure \ref{fig10a} and \ref{fig10b} pictorially represents the structured grid on base plane and symmetry plane of the waverider.

\subsection{Solver}
Freestream condition is set to Mach 5.5 for an altitude of 21 km in this study. The atmospheric pressure and temperature at said altitude are 467.76Pa and 217.65K. The reference length and area are the length of waverider and its planform area. The Reynold's number estimated for reference length of 0.2456m is 1.11x$10^6$. Hence, density based implicitly coupled Reynolds Averaged Navier Stokes (RANS) equations are used to numerically simulate steady-state high-speed flow in a commercial software, ANSYS Fluent.  \newline
\subsubsection{Turbulence modelling}
Spalart Allmaras (SA) single equation eddy-viscosity turbulence model developed by Spalart et al. \cite{Spalart1992} is widely used in aerospace applications. Validity of SA model for high speed flows is examined by Paciorri et al. \cite{Paciorri1997} for hollow cylinder-flare and hyerboloid flare configurations. The pressure measurements of the two models at Mach 5 and Mach 6.8 respectively were found to be in good agreement with experimental data. Roy et al. \cite{Roy2006} in an exhaustive survey consolidates the results of various turbulence models used for axi-symmetric models in high speed flows. SA performed well in flowfields involving attached flows and wakes.
\subsubsection{Solver setting}
\par Spalart Allmaras turbulence model with default coefficient settings as given by Matsson \cite{Matsson2021} calculates the viscous flowfield solution. Second order spatially accurate upwind scheme applies the Roe-FDS splitting method to fluxes and least squares cell method computes the gradients. Freestream air is calorically perfect with viscosity coefficient evaluated by Sutherland’s law. Convergence of the solution is assumed to be achieved when the residual (RMS) falls below $10^{-4}$

\subsubsection{Boundary conditions}
\begin{figure}[htb]
\centering
\includegraphics[width=0.7\textwidth]{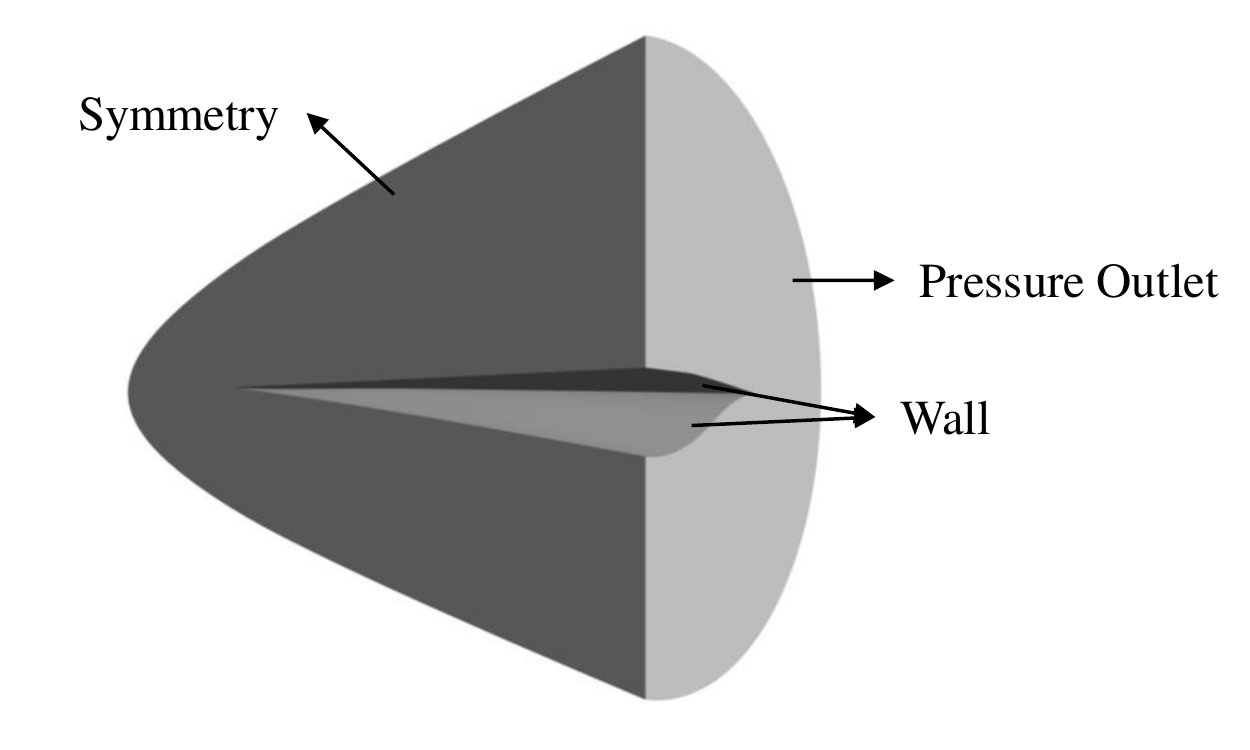}
\caption{Boundary conditions imposed on surfaces}
\label{fig11}
\end{figure}
The boundary conditions of the flowfield is schematically presented in Figure \ref{fig11}. While the farfield is conditioned to freestream flow, no-slip wall boundary condition is imposed on the model surface. The base plane is the pressure outlet with non-reflection boundary conditioned to pressure at infinity. y+ of 30 is chosen for the near wall mesh to maintain accuracy of the solution.

\subsection{Grid independence study}
Three structured grids are compared to check the accuracy of the solution. Table \ref{Table01} lists the grid sizes and the corresponding L/D obtained through RANS simulation. Coarse grid of 0.71M cells gives the maximum error of 0.1742\% relative to the fine grid of 2.68M cells. Hence the grid with 1.13M cells is employed for this work.

\begin{table}[htb]
\centering
\begin{tabular}{|c|c|c|}
  \hline
  Grid size ($10^6$) & L/D & \% Error \\ 
  \hline\hline
  0.71 & 4.0170 & 0.1740 \\ 
  \hline
  1.13 & 4.0262& 0.0547 \\ 
  \hline
  2.68 & 4.0240 & -\\
  \hline
\end{tabular}
\caption{}
\label{Table01}
\end{table}

\section{Results and discussion}\label{R&D}
\subsection{ANN Training and Hyperparameter Optimization}
\begin{figure}[htb]
\centering
\begin{minipage}{.48\textwidth}
\centering
\includegraphics[width=\linewidth]{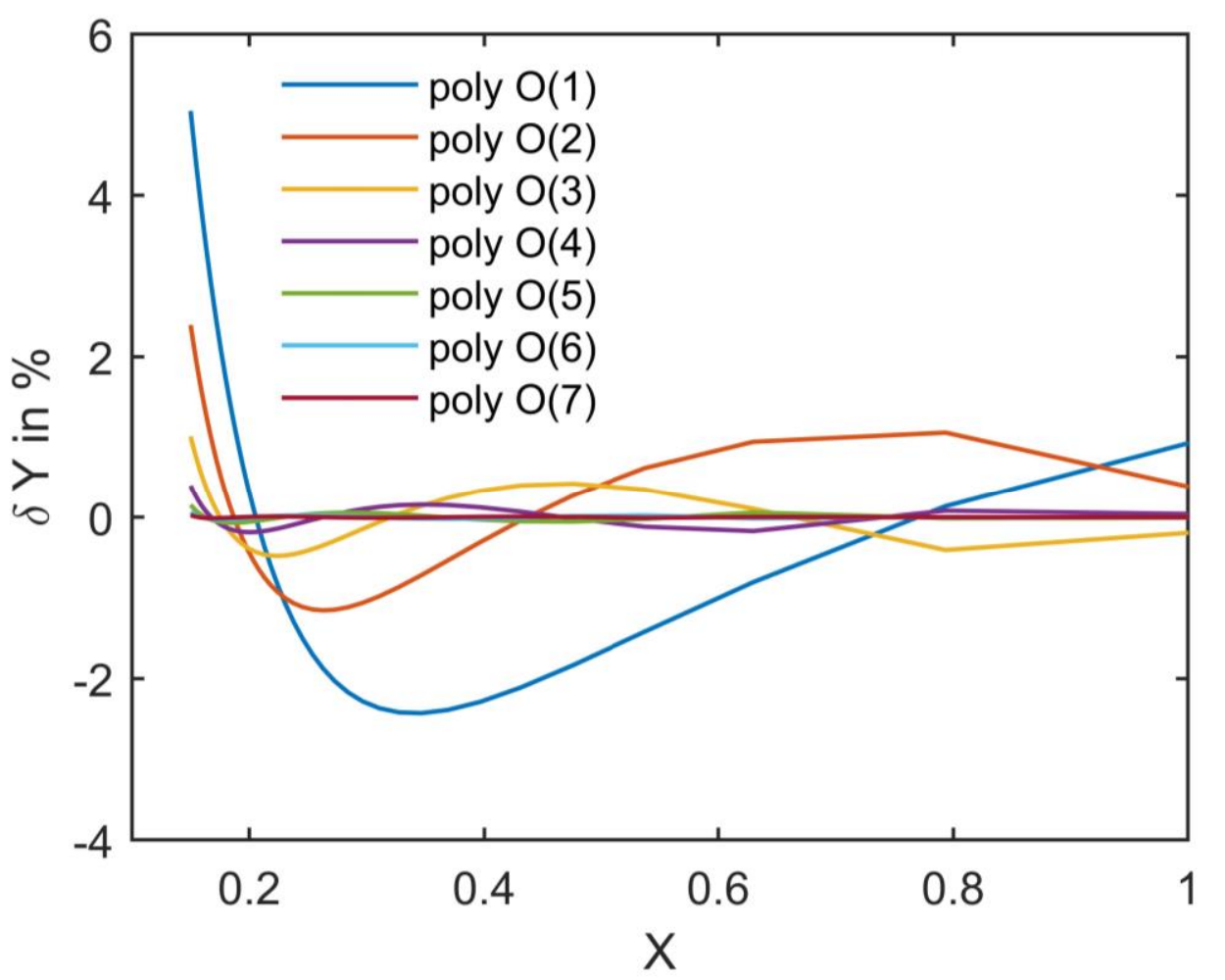}
\caption{\% error in Y of a streamline}
\label{fig12}
\end{minipage}
\hfill
\begin{minipage}{.48\textwidth}
\centering
\includegraphics[width=\linewidth]{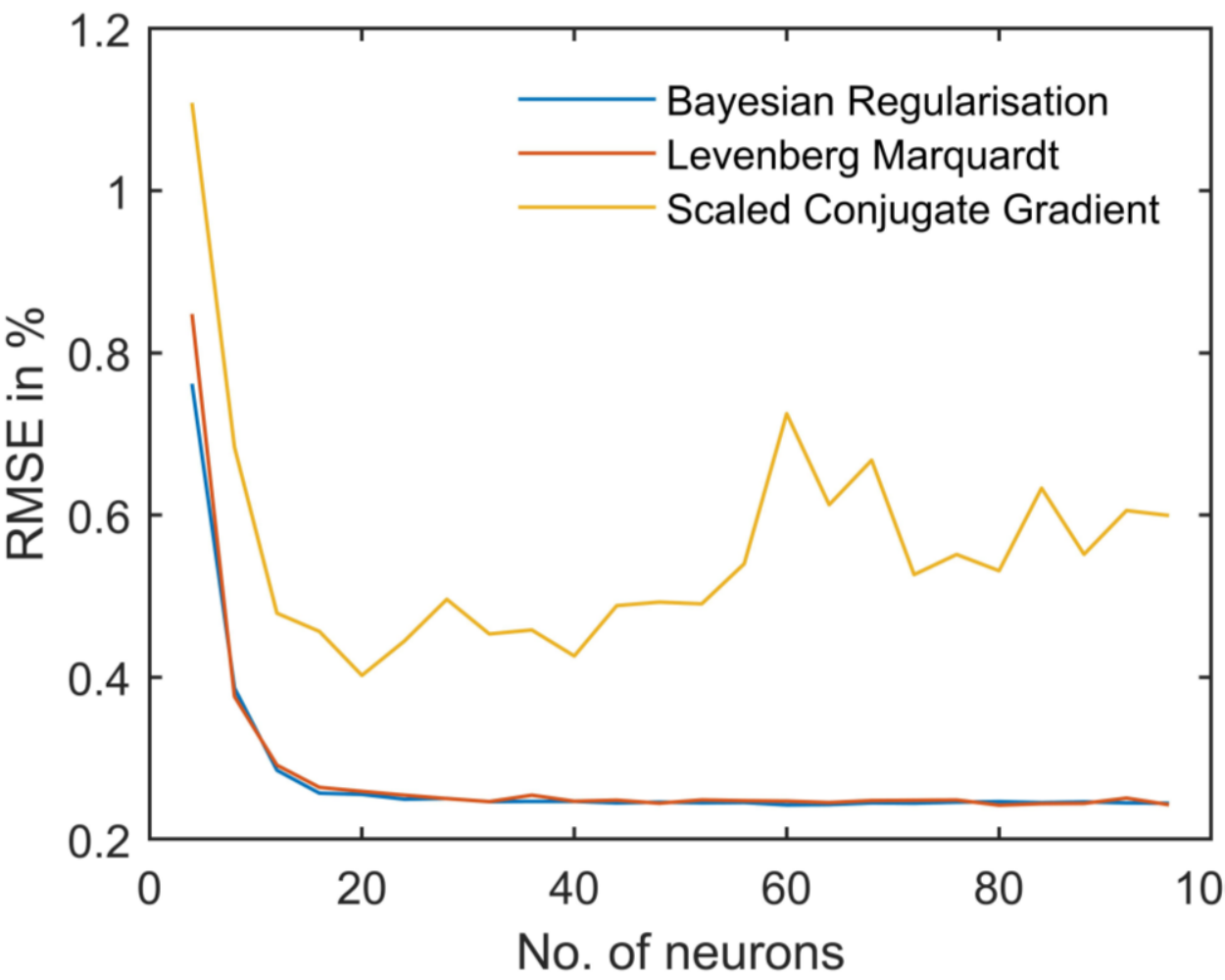}
\caption{RMSE vs. number of neurons}
\label{fig13}
\end{minipage}
\end{figure}

The data set created to train the ANN model is generated by solving TM equations for $M$ varying from 5 to 12 in steps of 0.25 and $\beta$ corresponding to cone angle varying from 5 to 45 $^\circ$ in steps of 5$^\circ$. Streamlines are chosen from $X_s$=0.1 to 1 along the length of the cone. The streamlines so obtained are curve-fitted to polynomials ranging from degree 1 to 7. Figure \ref{fig12} shows the error in Y for a streamline at $X_s$=0.1 at $M$ 5.5 and cone angle of 12$^\circ$. The maximum percentage error in Y for the given range of $(M,\beta, X_s)$ is found to be less than 8\% (highest error for first order polynomial) with the maximum lying at the smallest values of $(M,\beta, X_s)$ for all the polynomials.  The error is found to decrease with increasing degree of the polynomial. However, higher degree polynomials involve computation of many more coefficients, and hence to strike a balance the least degree of polynomial with a sufficiently accurate prediction is chosen. The percentage error in Y for third order polynomial is less than 2\% for all the values of inputs $(M,\beta, X_s)$. Hence to maintain minimum number of outputs without losing accuracy of the solution, third order polynomial is chosen in this study. 

\par
\begin{figure}[htb]
\centering
\includegraphics[width=0.75\textwidth]{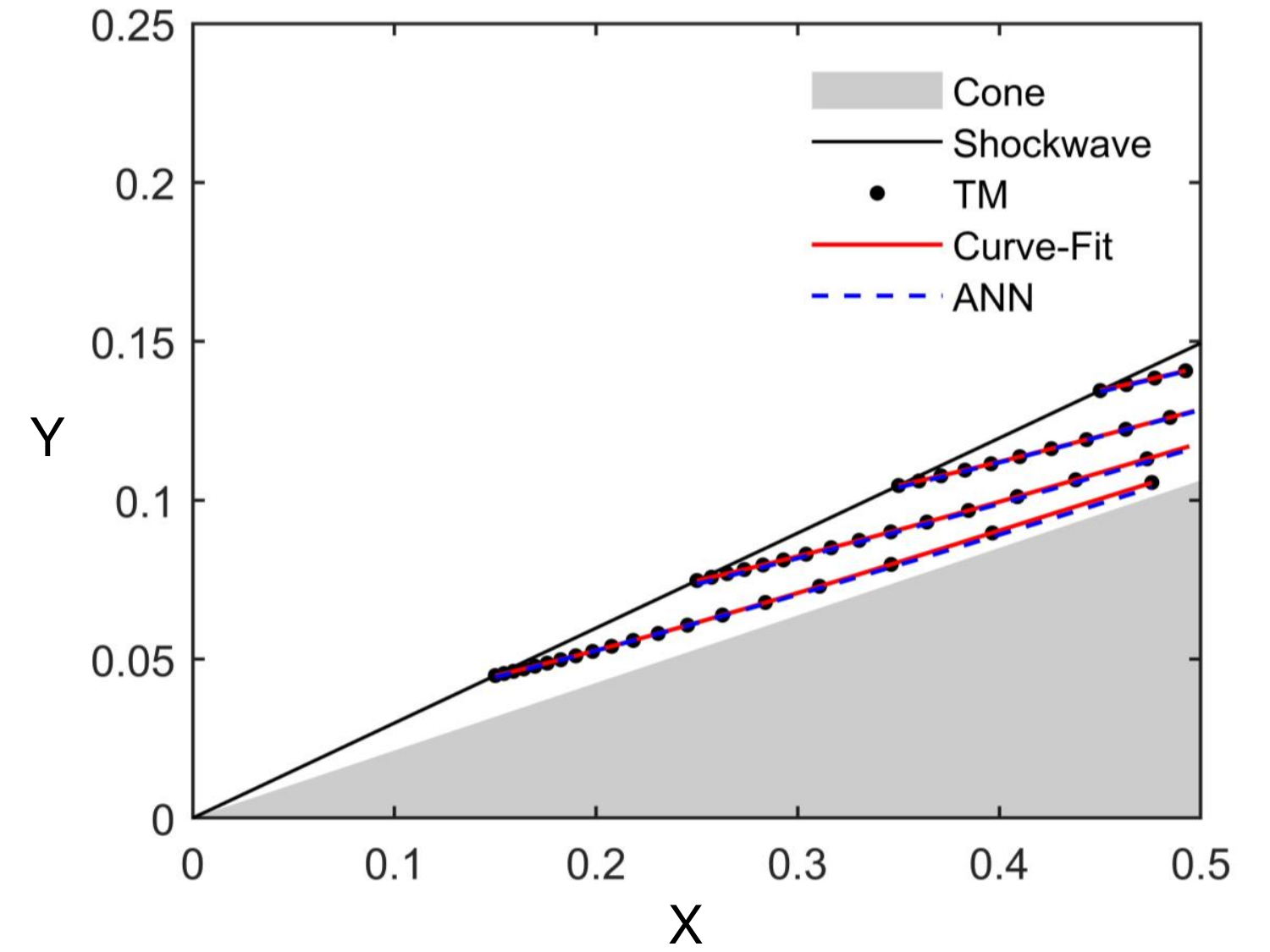}
\caption{A schematic of the streamlines}
\label{fig14}
\end{figure}

 Figure \ref{fig13} shows the influence of hyper-parameters on performance of the ANN model. It is observed that the RMSE (Root Mean Squared Error) for Bayesian Regularisation and Lavenberg Marquardt model stabilises with increase in number of neurons. For 32 neurons in a layer, the RMSE of the models mentioned are 0.2465\% and 0.2467\% respectively. Improvement in these value for 2 hidden layers is found to be minimal and hence single layered 32 neurons network with Bayesian Regularisation model is chosen. Figure \ref{fig14} pictorially depicts the streamlines obtained from ANN model in comparison with the standard and curve fitted streamlines. All three cases lie in close agreement to each other. 

\subsection{Computational Speed}
In order to indicate the effectiveness of the ANN model with respect to time consumption, the time taken to obtain the coordinates of 1 and 20 streamlines are tabulated in Table \ref{Table02}. A system with 11$^{th}$ Gen Intel Core i7 processor is used to give the estimate of time for both the cases. Evidently, the ANN model provides the streamline coordinates approximately 50 times faster and 20 times faster to generate 1 and 20 streamlines respectively when compared to the standard procedure to extract streamline coordinates explained in Section \ref{design} 

\begin{table}[htb]
\centering
\begin{tabular}{|c|c|c|}
  \hline
  No. of Streamlines  & \multicolumn{2}{|c|}{Time (s)}\\ \cline{2-3}
  & Standard & ANN \\
  \hline
  \hline
  1 & 0.3777 & 0.0073 \\ 
  \hline
  20 & 0.3726 & 0.0191 \\ 
  \hline
\end{tabular}
\caption{Time estimation}
\label{Table02}
\end{table}
\subsection{Comparisons of Waverider Generation Technique}
\par
\begin{figure}[htb]
\centering
\includegraphics[width=0.75\textwidth]{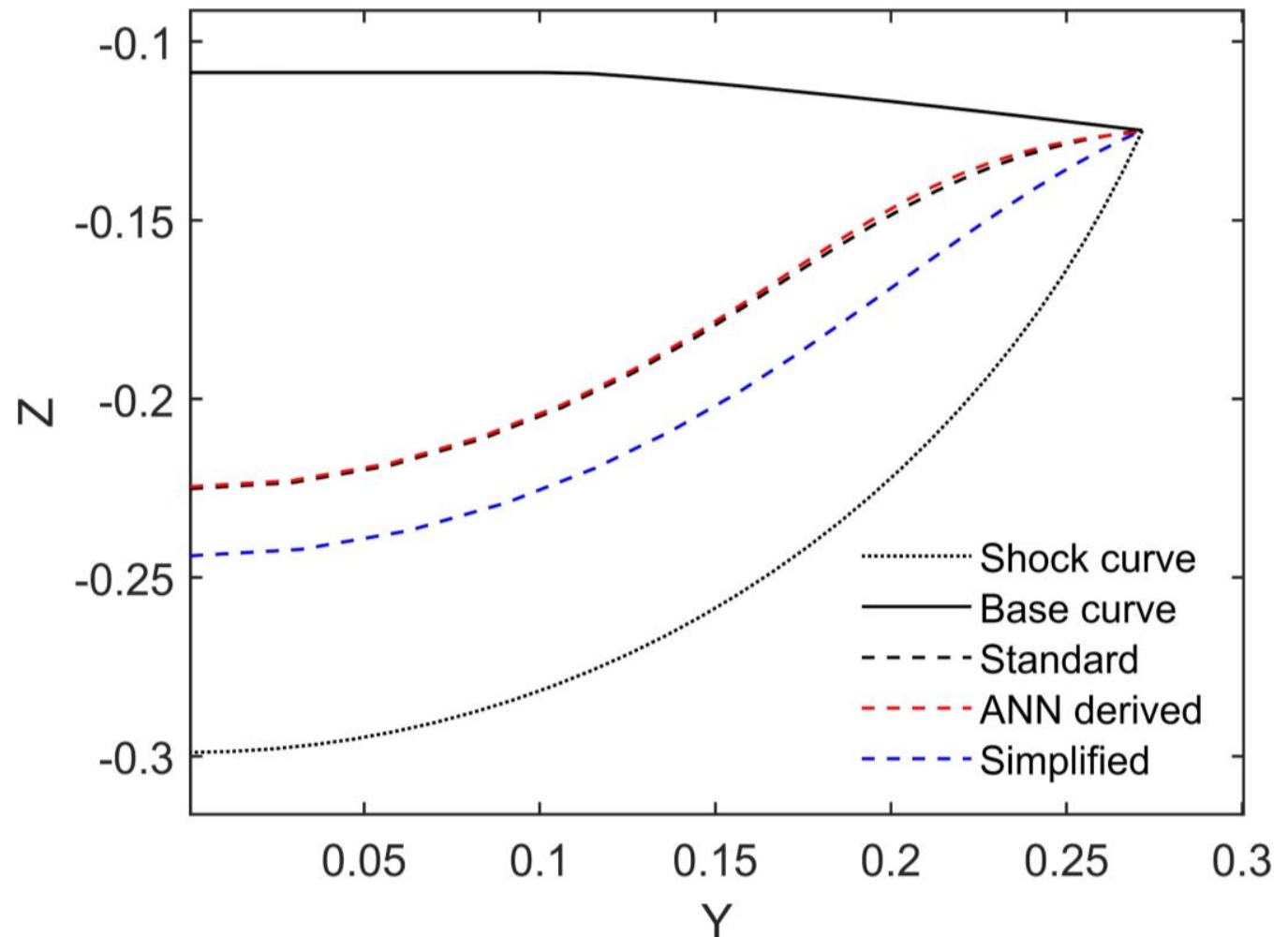}
\caption{Compression curve on base plane of the three models}
\label{fig15}
\end{figure}

 Three waverider configurations with the same base curve specification are designed based on the different streamline tracing strategy mentioned in Section \ref{method}. The standard waverider follows TM solution, the simplified waverider considers geometrical relations while the final waverider is designed by the novel ANN derived streamlines. The compression curve on the base plane for these three models are plotted in Figure \ref{fig15}. Maximum percentage error with respect to the standard approach in the coordinates of ANN derived model is is 0.68\% and that of the simplified model is 8.47\%.

\par The simplified conical waverider generation technique was introduced to reduce the computational complexity of the standard streamline tracing technique. In the simplified technique the streamlines are taken parallel to the conical surface. However, it is well known that the conical flow streamlines are not parallel to the conical surface in the near field and hence this discrepancy leads to wide differences with the standard approach. 
\par The ANN model is trained to follow the standard streamline tracing technique with high degree of accuracy. Instead of solving the streamline differential equations at every point the ANN model directly yields the equation of the streamline thus being more computationally effective with better accuracy.

\begin{figure}[htb]
  \begin{subfigure}[t]{0.32\textwidth}
    \includegraphics[width=\textwidth]{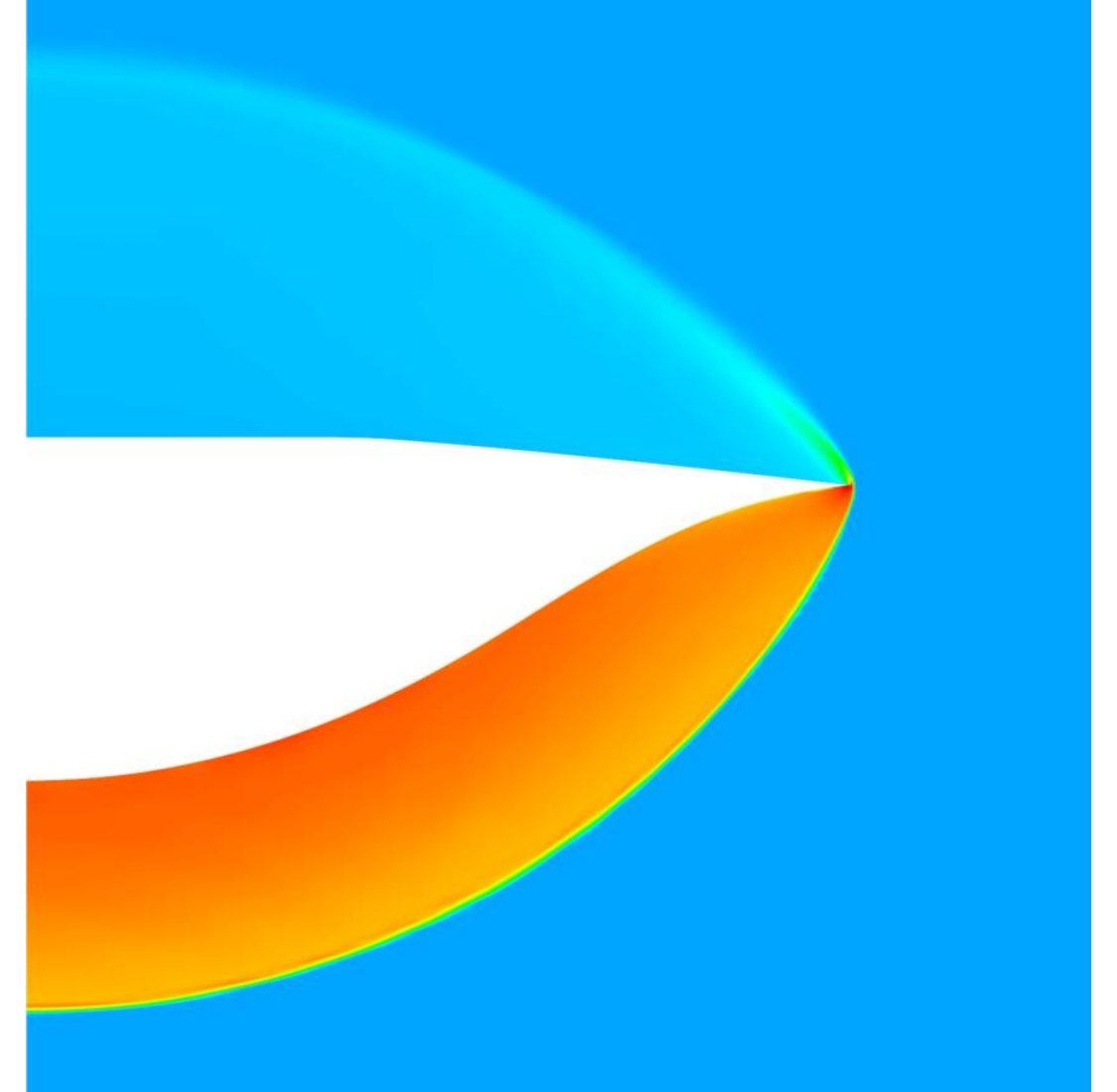}
    \caption{Standard}
    \label{fig16a}
  \end{subfigure}
    \hfill
  \begin{subfigure}[t]{0.32\textwidth}
    \includegraphics[width=\textwidth]{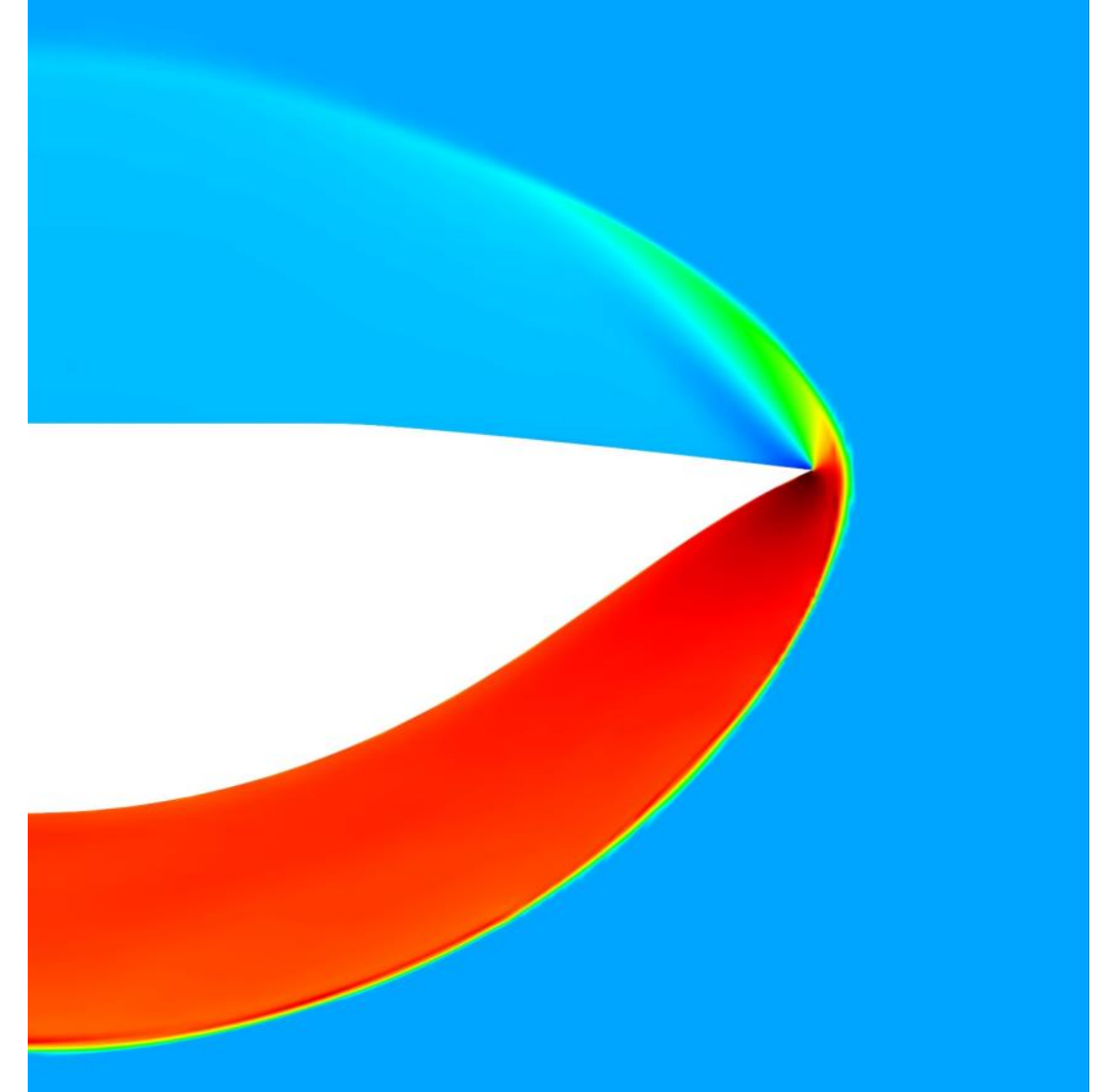}
    \caption{Simplified}
    \label{fig16b}
  \end{subfigure}
    \hfill
  \begin{subfigure}[t]{0.32\textwidth}   
  \includegraphics[width=\textwidth]{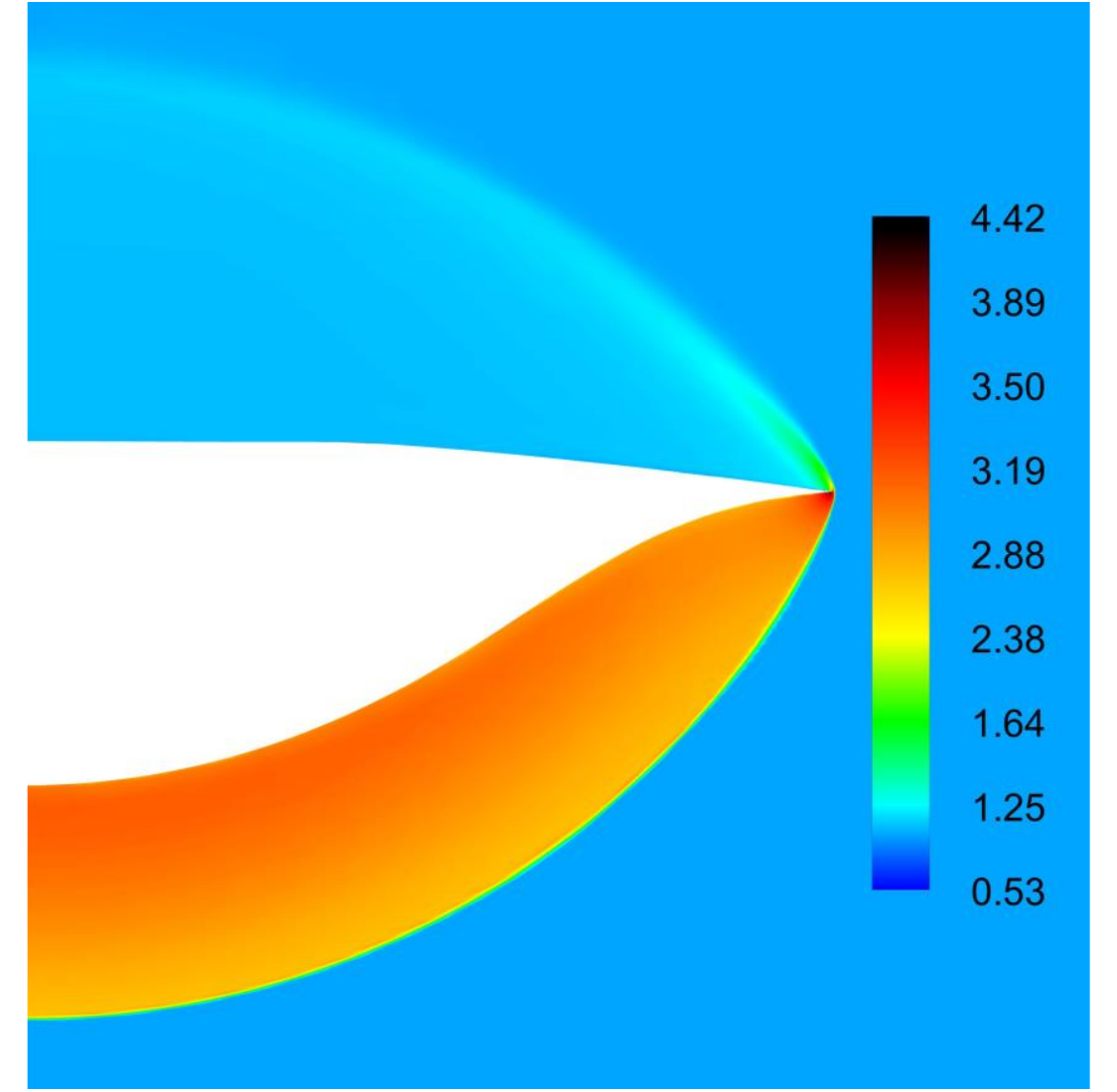}
   \caption{ANN derived}
   \label{fig16c}
  \end{subfigure}
  \caption{Pressure plot on Base plane of the waveriders}
  \label{fig16}
\end{figure}

Figure \ref{fig16} depicts the pressure plot on the base plane of the three comparative models with respect to freestream pressure. The shock angles are slightly higher than the angle predicted by inviscid TM solution due to the presence of boundary layer. The contours appear similar in case of Figure \ref{fig16a} and \ref{fig16c} while relatively high pressure is observed between the compression surface and shock of the simplified model in Figure \ref{fig16b}. This high pressure can be attributed to the increased shock angle perceived due to lack of curvature in the streamlines of the simplified model. Significant spillage is noticed at the tip of simplified model. The spillage is dramatically lower in the ANN derived model relative to the standard model. The effect of this spillage is evident through drop in aerodynamic efficiency of the waveriders. L/D of the standard and ANN derived model are noticeably close as shown in Table \ref{Table03} while that of the simplified model drops by 12.52\%. 

In addition to L/D, Table \ref{Table03} also presents the volumetric efficiencies of the models. If V is the internal volume and S the planform area of the waverider then its volumetric efficiency $\eta$ is given by $V^{2/3}/S$. Planform area of all the three models remain the same and is found to be 0.2089 $m^2$. While difference in $\eta$ of the standard and ANN derived model is negligible, higher value for the simplified model is owed to increase in its internal volume on considering streamlines parallel to the cone.

\begin{table}[htb]
\centering
\begin{tabular}{|c|c|c|}
  \hline
  Models & L/D & $\eta$\\ 
  \hline
  \hline
  Standard & 4.0262 & 0.2027\\ 
  \hline
  Simplified & 3.5219 & 0.2236\\ 
  \hline
  ANN derived & 3.9749 & 0.2023\\ 
  \hline
 \end{tabular}
\caption{L/D and volumetric efficiencies of the waveriders}
\label{Table03}
\end{table}

\section{Conclusion}
Streamline tracing from axis-symmetric conical flow is an indispendable tool for generation of hypersonic waveriders and intakes critical to the development of hypersonic flight systems. The standard streamline tracing technique is computationally cumbersome involving sequential solution of the Taylor-Maccoll equations and then the differential equations for the streamlines. In the case of waverider geometry generation, a certain simplification was achieved through the simplified method, but the coordinates of the waverider geometry had significant differences due to the assumption of streamlines being parallel to the conical surface. 
\par We have developed a novel ANN-based streamline tracing technique which is both computationally efficient and highly accurate with respect to the standard approach. The ANN model is trained to output the coefficients of a third order polynomial fit to the streamlines given the upstream Mach number, shock angle and the streamline originating location. The new ANN-based approach yields streamline coordinates within 0.68\% difference to the standard approach and is about 20 times faster. The waverider generated from the ANN-based approach closely resembles the one from the standard approach and has much smaller spillage at the tips.

%% If you ha ave bibdatabase file and want bibtex to generate the
\clearpage
 \bibliographystyle{elsarticle-num} 
 \bibliography{BibDatabaseWaverider}





% \end{thebibliography}
\end{document}